\documentclass[a4paper, journal]{IEEEtran}
\usepackage{cite}
\usepackage{latexsym}
\usepackage{textgreek}
\usepackage{etoolbox}
\makeatletter
\patchcmd{\@makecaption}
  {\scshape}
  {}
  {}
  {}
\makeatletter
\patchcmd{\@makecaption}
  {\\}
  {.\ }
  {}
  {}
\makeatother
\def\tablename{Table}
\usepackage{tcolorbox}
\usepackage{color}
\usepackage{fontawesome}
\usepackage{amsthm}
\usepackage{amsxtra}
\usepackage[mathscr]{euscript}
\usepackage{graphicx}
\graphicspath{{./fig/}}
\usepackage{multirow}
\usepackage{pifont}

\usepackage{rotating, graphicx}
\usepackage{array,multirow,graphicx}
\usepackage{amsmath}
\usepackage{tikz}
\usepackage{amssymb}
\usepackage{lipsum}
\usepackage{lipsum, multicol}
\usepackage{adjustbox}
\usepackage{algorithm}
\usepackage{algpseudocode}
\usepackage{upgreek}
\usepackage{mathtools}
\usepackage{float}
\usepackage{xcolor}
\usepackage[caption = false]{subfig}
\usepackage{acronym}
\usepackage{orcidlink}
\usepackage{cuted} % For spanning content across columns
\usepackage{fancyhdr}
\hypersetup{hidelinks}
\usepackage{colortbl,booktabs}
\usepackage{svg}
\usepackage{supertabular}
\usepackage{comment}
\usepackage{alltt}
\usepackage{enumitem}
\usepackage{mathtools}
\usepackage{float}
\usepackage[caption = false]{subfig}
\usepackage{threeparttable}
\usepackage[utf8]{inputenc}
\usepackage[english]{babel}

\makeatother
\usepackage{eso-pic}
\newcommand\AtPageUpperMyright[1]{\AtPageUpperLeft{
 \put(\LenToUnit{0.05\paperwidth},\LenToUnit{-1cm}){
     \parbox{1\textwidth}{\raggedleft\fontsize{9}{11}\selectfont #1}}
 }}
\newcommand{\conf}[1]{
\AddToShipoutPictureBG*{
\AtPageUpperMyright{#1}
}
}
\begin{document}

\title{The Structure of Service Level Agreement of Slice-based 5G Network
\vspace{-0.40mm}}

\conf{This article has been accepted for publication in the IEEE PIMRC 2018, 09-12 September 2018, Bologna, Italy.} %Header itself.

\author{Mohammad Asif Habibi\orcidlink{0000-0001-9874-0047}, 
Bin Han\orcidlink{0000-0003-2086-2487}, 
Meysam Nasimi\orcidlink{0000-0001-7743-5246}, 
and Hans D. Schotten\orcidlink{0000-0001-5005-3635}
\vspace{-11.8mm}
\thanks{This work has been performed with the financial supports of the H2020-MSCA-ITN-2015 project 5GAuRA.}
\thanks{Mohammad Asif Habibi\orcidlink{0000-0001-9874-0047}, Bin Han\orcidlink{0000-0003-2086-2487}, Meysam Nasimi\orcidlink{0000-0001-7743-5246}, and Hans D. Schotten\orcidlink{0000-0001-5005-3635} are with the Division of Wireless Communications and Radio Navigation (WiCoN), Department of Electrical and Computer Engineering (EIT), University of Kaiserslautern (RPTU), $\mathrm{67663}$ Kaiserslautern, Germany. Hans D. Schotten\orcidlink{0000-0001-5005-3635} is also affiliated with the Intelligent Networks (IN) Research Department, German Research Center for Artificial Intelligence (DFKI), $\mathrm{67663}$ Kaiserslautern, Germany. The corresponding author is Mohammad Asif Habibi\orcidlink{0000-0001-9874-0047} (\texttt{asif@eit.uni-kl.de}).} }

\maketitle
\begin{abstract}
Network slicing is considered one of the key enablers of fifth-generation (5G) communication systems. Traditional telecommunication networks have provided various services to diverse customers through a single shared network infrastructure. In contrast, with the deployment of network slicing, operators can partition the network into multiple slices, each with its own configuration and quality of service (QoS) requirements. Numerous applications across different industries require dedicated slices, each tailored with specific functions and features. These applications create new business opportunities that necessitate new business models; consequently, each slice requires an individual service-level agreement (SLA). In this paper, we propose a comprehensive end-to-end SLA structure between the tenant and the operator in a slice-based 5G network, aiming to balance the interests of both parties. The proposed SLA framework defines the reliability, availability, and performance of delivered telecommunication services to ensure that the right information reaches the right destination at the right time in a safe and secure manner. Furthermore, we discuss key metrics for slice-based network SLAs, including throughput, penalties, cost, revenue, profit, and QoS-related metrics, which are critical considerations during SLA formulation and negotiation.
\end{abstract}

\IEEEoverridecommandlockouts
\begin{keywords}
5G, Metrics, Network Slicing, Operator, Service Level Agreement, Slice, Tenant, Wireless Communication
\vspace{-2.5mm}
\end{keywords}

\IEEEpeerreviewmaketitle

\section{Introduction}\label{sec:Introduction}
\vspace{-1.5mm}
\IEEEPARstart{T}{he} fifth-generation (5G) communication system is expected to fulfill diverse service requirements across many aspects of human life. It will enable various services for different vertical industries, such as automotive, logistics, healthcare, manufacturing, and agriculture. In addition, 5G is expected to support enhanced service experiences, including ultra high definition (UHD) video, online gaming, augmented and virtual reality, and cloud desktop services, even in scenarios characterized by ultra-high traffic density, high mobility, extremely high connection density, and wide coverage.

Most existing communication networks are monolithic, where a \textit{one-size-fits-all} architectural solution is used to provide services. However, to support various types of 5G applications and meet the diverse service requirements expected in the future, such a monolithic architecture is no longer sufficient. Consequently, the concept of \textit{network slicing} has emerged, whereby a network operator logically partitions its network into multiple virtual networks, referred to as \textit{network slices} \cite{asif2017network}. All network slices are deployed over the same physical infrastructure, while each slice maintains its own characteristics, such as quality of service (QoS) requirements, resource management mechanisms, architecture, and configuration. Network slicing enables operators to partition networks in a structured, elastic, scalable, and automated manner, thereby reducing operational costs, lowering energy consumption, and simplifying network management.

As briefly discussed earlier, each use case in a 5G communication system requires a dedicated network slice consisting of independent functions, requirements, and characteristics. For instance, a slice may be dedicated to critical machine-type communication (C-MTC), such as remote surgery, which is typically characterized by high reliability, ultra-low latency, and high throughput. Another network slice may be designed to support water meter reading, which requires a simple radio access procedure, small payload volumes, and low mobility. Furthermore, enhanced mobile broadband (eMBB) services may require a separate slice characterized by large bandwidth to support high data rate services such as HD video streaming. All these and other types of slices create new business opportunities that require new business models.

In legacy networks, the requirements of different service types are largely similar; therefore, most service level agreements (SLAs) between an operator and a tenant contain the same set of metrics. However, in 5G, each slice requires an individual SLA with unique elements, metrics, and structure compared to the SLAs of other slices within the same network.

As its name implies, an SLA is a formal agreement between a network operator and a tenant, or between network operators, in which the level of delivered service is precisely defined. According to the International Telecommunication Union (ITU), \textit{``the SLA is a formal agreement between two or more entities that is reached after a negotiating activity with the scope to assess service characteristics, responsibilities and priorities of every part"} \cite{ITUSLADefiniation}. An SLA establishes a common understanding of a service and its relevant aspects, such as performance, availability, and responsibilities.

Each SLA includes a number of specific elements, referred to as \textit{metrics}. These metrics describe the level and volume of communication services and are used to measure the performance characteristics of service components. An SLA also includes technical, economic, and legal statements in order to cover all aspects that must be agreed upon between the network operator and the tenant.

To efficiently measure service performance and accurately describe the service level, SLA management should be automated to ensure accountability under various network and business conditions as well as diverse user patterns across different network slices. The automated management of slice-based SLAs can be achieved through network programmability, virtualization, and control functions.

Recently, SLAs in telecommunication networks have been extensively studied. The International Telecommunication Union (ITU) proposed a generic structure for an SLA in a multi-service-provider telecommunication environment in Recommendation E.860 \cite{ITUSLADefiniation}. The proposed SLA defines QoS-related terms and further describes the entire procedure for establishing an end-to-end (E2E) SLA.

The European Telecommunications Standards Institute (ETSI) has also conducted numerous studies on SLAs, which are reported in \cite{ETSI_SLA}, \cite{ETSI_SLA001}, and \cite{ETSI_SLA002}. Reference \cite{ETSI_SLA} explores two main aspects of SLAs, namely the development phases and the SLA template, and further discusses the contents, technical features, QoS metrics and commitments, charging and billing, and reporting mechanisms of an SLA. Reference \cite{ETSI_SLA001} investigates the SLA life cycle and penalty mechanisms, while \cite{ETSI_SLA002} studies user demands and the various services provided to the tenants by a service provider.

Moreover, an E2E structure of QoS-oriented SLAs and a framework for real-time SLA management in multi-service packet networks are investigated in \cite{1003036}. The authors presented a monitoring scheme capable of generating revenue through admission flows and calculating penalties when flows are lost. However, to the best of our knowledge, no study to date has specifically explored the SLA between a tenant and an operator in a slice-based 5G network.

In this paper, we propose an E2E SLA structure for a slice-based 5G communication network. We further discuss the metrics of the proposed SLA that should be considered by both the operator and the tenant during the agreement process. The rest of the paper is organized as follows. The concept and structure of the proposed SLA are thoroughly discussed in Sec.~\ref{sec:ProposedSLA}. Subsequently, the metrics of the proposed SLA are analyzed in Sec.~\ref{sec:ElementsOfSLA}. Finally, Sec.~\ref{sec:Conclusion} summarizes the main conclusions and outlines directions for future research.
\vspace{-1.5mm}

\section{The Structure of Proposed SLA} \label{sec:ProposedSLA}
\vspace{-1.5mm}
In this section, we introduce and thoroughly describe the E2E structure of the proposed slice-based SLA between a tenant and an operator in a 5G communication system. Moreover, we discuss two types of slice-based SLAs, namely \textit{Static SLA} and \textit{Dynamic SLA}, which help simplify the operation of different categories of services across various types of slices within the same 5G communication network.

A static SLA is a predefined agreement in which all metrics, the quality of assured service, as well as legal and financial terms are specified in advance between the tenant and the operator. Once the static SLA becomes active, the service operates according to the agreed parameters, and neither party can modify the SLA terms -- such as increasing throughput or decreasing latency -- during its lifetime.

In contrast, in a dynamic SLA, the values of certain metrics may change dynamically according to the tenant’s requirements. For example, the tenant of a low-latency slice may pay according to the amount of allocated bandwidth; the higher the bandwidth allocation, the higher the payment required. Alternatively, a tenant may require full control of the slice and guaranteed extremely low-latency service during a remote surgery session. Once the surgery is completed, the network slice may cease providing the service.

The entire lifecycle of a slice-based SLA consists of three phases: the creation phase, the operation phase, and the termination phase. In the creation phase, the tenant selects an operator capable of fulfilling its requirements. Subsequently, both parties agree upon and establish the SLA, after which the service begins operating over the slice.

During the operation phase, the service remains under continuous maintenance and monitoring by both parties. In the event of an SLA violation, the corresponding \textit{penalty} mechanisms are enforced.

In the termination phase, which may be triggered either by a violation of the agreement or by the expiration of the contract, the slice ceases to provide services and the SLA is terminated. Once the decision is made to remove the slice and terminate the SLA, it is recommended that all information associated with service configuration, tenant service requirements, and service maintenance be removed from the system. However, some tenants or operators may prefer to archive service-related information for a certain period of time.

\begin{figure}
  \centering
  \includegraphics[width=0.5\textwidth]{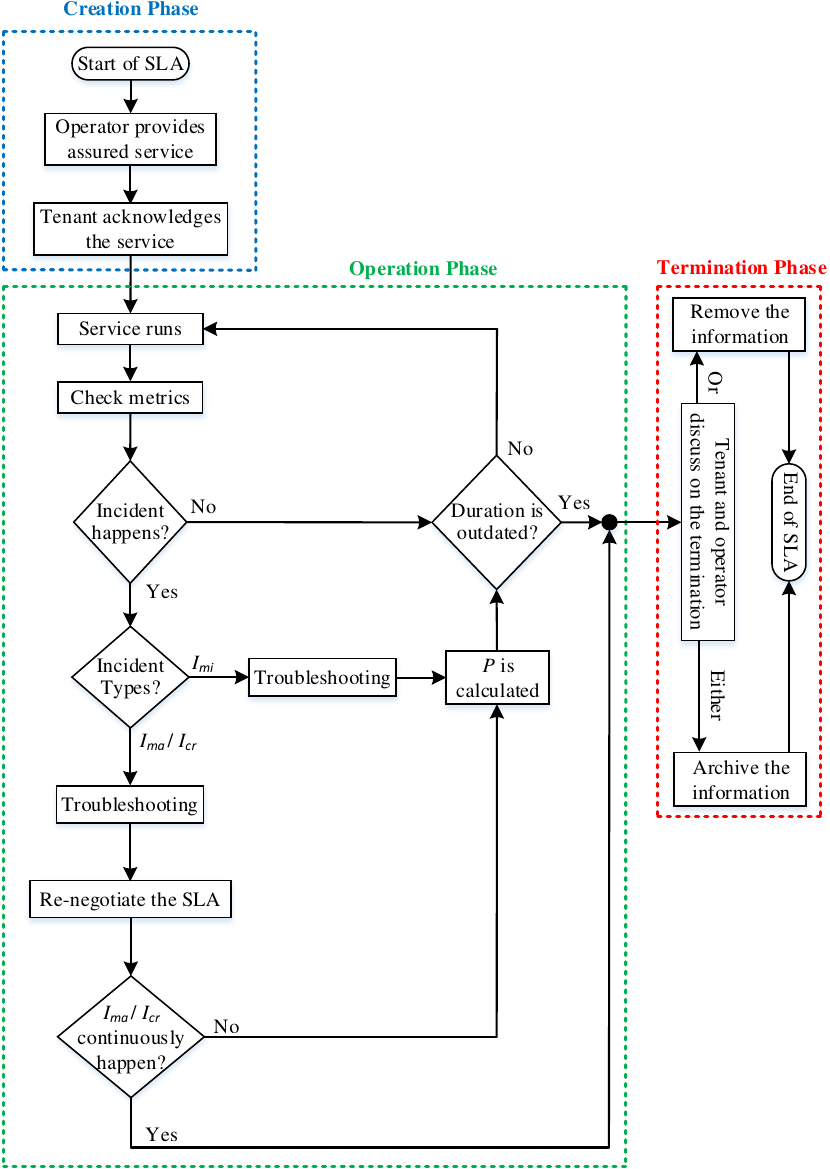}
  \caption{An E2E structure of proposed SLA in a slicing-based network}
  \label{fig:structure_of_SLA}
\end{figure}

The detailed procedure of the proposed SLA is illustrated in Fig.~\ref{fig:structure_of_SLA}. In the creation phase, the tenant and the operator agree on all terms and conditions of the agreement. Within this agreement, the tenant is guaranteed a certain level of QoS for a specified period of time, referred to as the \textit{SLA lifetime}. Once the agreement is finalized, both parties sign the relevant documents, and the SLA is formally established.

During the operation phase, the operator provides and maintains services for the tenant through a dedicated slice, which is acknowledged by the tenant. Meanwhile, a set of QoS metrics associated with the slice service, such as security, power consumption, throughput, and latency, are continuously monitored in real time. The monitoring functionality of the slice should be accessible to both parties to ensure proper service configuration, management, and maintenance.

In the context of slice-based SLAs, incidents that may occur within a slice can be categorized into three levels: minor incidents ($I_{\mathrm{mi}}$), major incidents ($I_{\mathrm{ma}}$), and critical incidents ($I_{\mathrm{cr}}$). A minor incident ($I_{\mathrm{mi}}$) indicates a noncritical condition within the slice that, if left unaddressed, may lead to service interruption or performance degradation. Although such incidents typically do not disrupt the entire slice, they may degrade a limited portion of the service.

A major incident ($I_{\mathrm{ma}}$) requires immediate response, as it poses a serious risk to the integrity of the network, for example during traffic overload situations. A critical incident ($I_{\mathrm{cr}}$) represents the most severe condition within the slice and is typically caused by failures of hardware components. Once an incident occurs, all monitoring metrics should be automatically examined for troubleshooting and for evaluating potential SLA breaches, as well as for determining the type of incident. If $I_{\mathrm{mi}}$ occurs within the slice, it should be resolved as soon as possible. After resolving $I_{\mathrm{mi}}$, a penalty $P$ may be calculated according to the source and severity of the incident, which the operator is required to pay to the tenant. In the case of $I_{\mathrm{mi}}$, we recommend that the operator and the tenant agree on specific threshold values for each monitoring metric associated with penalties. In this context, the tenant does not impose a penalty on the operator if the incident can be resolved without violating any of the predefined thresholds. Otherwise, the tenant imposes a penalty $P$ on the operator and formally notifies the operator to resolve the incident promptly and ensure the quality of the agreed services.
\vspace{-1mm}
Let $P$ denote penalty related to a single breach event, and let $P_{\mathrm{tot}}$ denote total penalty accumulated over an SLA observation window (or, equivalently, over the SLA lifetime when the observation window coincides with the SLA lifetime).
\vspace{-1mm}

Assuming that either $I_{\mathrm{ma}}$ or $I_{\mathrm{cr}}$ occurs, these incidents are typically more difficult to resolve. Therefore, we recommend that the SLA clearly specify how effectively and within what time frame $I_{\mathrm{ma}}$ and $I_{\mathrm{cr}}$ must be addressed. In the case of $I_{\mathrm{ma}}$ or $I_{\mathrm{cr}}$, both parties may renegotiate and further optimize the SLA in response to major or critical incidents, which can help them avoid further service interruptions. Moreover, a long-term tracking mechanism for the occurrences of $I_{\mathrm{ma}}$ and $I_{\mathrm{cr}}$ is introduced, allowing tenant to terminate the SLA before the end of its lifetime and switch to another qualified operator if such serious incidents occur repeatedly. Otherwise, the SLA remains valid until it expires, at which point the slice ceases to provide the service, and both the operator and the tenant finalize all outstanding matters, including financial and legal obligations, arising during the contractual period.

\section{The Metrics of the Proposed SLA} \label{sec:ElementsOfSLA}
\vspace{-1.5mm}
One of the primary purposes of an SLA is to define appropriate and realistic metrics for the services delivered by the operator to the tenant. These metrics should be continuously monitored in order to detect potential breaches of the agreement. In this section, we discuss several key concepts related to slice-based network SLAs, including service availability, penalties, cost, revenue, profit, and QoS-related metrics.
\vspace{-1.5mm}

\subsection{Service Availability}
\vspace{-1.5mm}
The measurement of service availability has a long history in the telecommunication industry. It is one of the most important SLA metrics for both the tenant and the operator, and it must be defined as clearly as possible to avoid any misunderstanding between the two parties. The International Organization for Standardization (ISO) defines availability as \textit{``the ability of a functional unit to be in a state to perform a required function under given conditions at a given instant of time or over a given time interval, assuming that the required external resources are provided"} \cite{ISODefiniation}.

In a simplified sense, availability refers to the successful delivery of service or data from point A to point B and is typically expressed as a dimensionless ratio in percentage form. The period during which a network or slice is unable to deliver service or data to the tenant is referred to as \textit{downtime}.

Let $T_{\mathrm{h}}>0$ denote the total observation window for the slice service, and let $T_{\mathrm{u}}$ represent the total unavailable time of that service within the same window, where $0 \le T_{\mathrm{u}} \le T_{\mathrm{h}}$. The service availability can then be expressed as \begin{equation}
A = \frac{T_{\mathrm{h}} - T_{\mathrm{u}}}{T_{\mathrm{h}}}
  = 1 - \frac{T_{\mathrm{u}}}{T_{\mathrm{h}}}.
\label{eq:service_availability}
\end{equation}

We categorize the service availability of a slice into three ranges: high availability (e.g., $\approx 100\%$), average availability (e.g., $\ge 99.5\%$), and low availability (e.g., $< 99\%$). This categorization helps both the operator and the tenant evaluate whether the measured performance of a slice meets, exceeds, or falls below the predefined availability levels within a given period of time. Both parties should formally agree on conditional guarantees. For example, if the average service availability of a slice during a specified period falls below $99\%$, the operator is required to pay a penalty to the tenant.
\vspace{-1.5mm}
\subsection{Penalty}
\vspace{-1.5mm}
Most network operators promise to guarantee a high level of network performance. However, these guarantees are not always fulfilled; therefore, it is advisable for both the operator and the tenant to predefine an appropriate penalty value in the SLA. This penalty may be imposed by the tenant when the operator fails to deliver the agreed level of service.

In the context of an SLA, limited levels of service incidents or temporary unavailability may be tolerated. However, any degradation beyond these predefined thresholds is considered unacceptable, and the operator should be penalized according to the terms of the agreement. In some cases, the tenant may attempt to maximize the penalty in order to encourage the operator to maintain the required service level. Conversely, the network operator may try to negotiate lower penalty values in case of service failure or include contractual terms that could reduce the level of guaranteed service.

Nevertheless, well-informed operators and tenants typically avoid such unfavorable conditions, as they may result in either excessive penalties or unacceptable service degradation. It is also worth noting that the term ``penalty,'' although widely used by both tenants and operators, is not legally precise. From a legal perspective, the term ``fee reduction'' is considered more appropriate to describe this concept \cite{Gartner2003}.

\begin{figure}%[!h]
  \centering
  \includegraphics[width=0.5\textwidth]{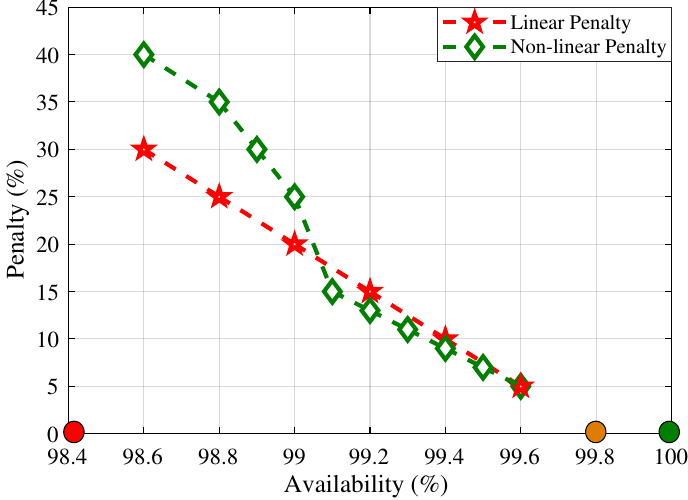}
  \caption{Linear penalty and non-linear penalty}
  \label{fig:Penalties}
\end{figure}

We divide penalties into two types: \textit{linear penalty} and \textit{non-linear penalty}. In a linear penalty, the tenant charges the operator a predefined penalty when the service availability falls below a specified threshold. As illustrated in Fig.~\ref{fig:Penalties}, we consider 100\% as the agreed availability, 99.8\% as the accepted availability, and 98.4\% as the terminated availability. Between the accepted and terminated availability levels, penalties are imposed according to a predefined linear rule.

Specifically, we assume that for each 0.2\% decrease in availability below the accepted level, the network operator is charged a penalty of 5\%. Based on these assumptions and the results presented in Fig.~\ref{fig:Penalties}, when availability decreases to 99.6\%, a 5\% penalty is imposed, while an availability of 99.4\% results in a 10\% penalty, and so on.

In contrast, in a non-linear penalty model, the operator and tenant agree on irregular penalty amounts corresponding to different availability levels. In this case, there is no fixed linear relationship between the level of availability and the amount of penalty. We assume that the operator incurs a 5\% penalty when availability falls 0.2\% below the accepted level. Subsequently, an additional 2\% penalty is imposed for each further 0.1\% decrease in availability until the level reaches 99.1\%. Furthermore, if availability falls below 99.1\%, a penalty of 10\% is applied until the availability reaches 99\%. Below 99\%, an additional 5\% penalty is imposed until the availability reaches the terminated threshold. According to these assumptions and the results shown in Fig.~\ref{fig:Penalties}, the penalty reaches 25\% when the availability drops to 99\%, and increases to 35\% when availability decreases to 98.8\%, and so forth.

It is worth noting that if the service availability falls below the predefined terminated availability level, the tenant may terminate the network slice and migrate to another network operator capable of providing the required QoS. Moreover, when comparing linear and non-linear penalty models, the primary difference lies in the correlation between the level of availability and the amount of penalty imposed.

The above-mentioned linear and non-linear penalty models do not address all scenarios that may arise in the SLA of a complex slice-based network. In slice-based networks, tenant requirements may vary from one slice to another; similarly, each slice may have its own quality of service (QoS) requirements. Therefore, it is necessary to further investigate additional dimensions of penalties, such as the importance of the time at which a service breach occurs, the total number of failures within a given SLA period, the duration of each failure, and the total duration of all failures during the SLA period. To address these aspects, the concept of penalties in the context of slice-based network SLAs must be further developed through appropriate mathematical formulations \cite{4476555}.
\vspace{-1.5mm}

\subsection{Penalty Formulas and Variable Definitions}
\vspace{-1.5mm}
In the proposed SLA workflow (Fig.~\ref{fig:structure_of_SLA}), whenever an incident results in an SLA breach, the corresponding penalty is calculated during the operation phase. Over an SLA observation window, these penalties accumulate to form the total penalty $P_{\mathrm{tot}}$ (i.e., $P_{\mathrm{tot}}$ represents the sum of penalties associated with all breach events within the observation window). The total penalty can be defined under different SLA policies based on (i) the number of breaches, (ii) the duration of each breach, and (iii) the time-varying importance of the service, either at the slice level or at the subcontract level.

\subsubsection{Per-breach Charge}
The simplest SLAs may define a constant penalty value that does not depend on either the duration of the breach or the time at which it occurs. Let $n$ denote the number of SLA breach events within the observation window, and let $V$ denote the penalty amount per breach event. Then
\begin{equation}
P_{\mathrm{evt}} = Vn,
\label{eq:pen_evt}
\end{equation}
\noindent where $n$ is the number of breach events within the observation window and $V$ is the penalty per breach event.

\subsubsection{Charge Dependent on Breach Duration}
The penalty can also be proportional to the total unavailability duration. Let $T_{\mathrm{u}}$ denote the total unavailability time within the observation window, and let $w$ denote the penalty rate per unit time (at the slice level). Then \begin{equation}
P_t = wT_{\mathrm{u}},
\label{eq:pen_time}
\end{equation}
\noindent where $T_{\mathrm{u}}$ is the total unavailability duration within the observation window and $w$ is the penalty rate per unit time of service unavailability.

\subsubsection{Charge Dependent on the Time of Occurrence (Slice Level)}
In some SLAs, the same outage may have different levels of importance depending on when it occurs (e.g., peak versus off-peak periods). If penalties are evaluated at the slice level (without distinguishing between subcontracts), let $\delta t$ denote the time interval between measurements, $t_j$ the $j$-th time instant, and $I(t_j)$ the time-varying importance weight (dimensionless, typically $0 < I(t_j) \le 1$). Let $T$ denote the end time of a failure period and $\Delta T$ its duration, and define $J := \Delta T / \delta t$. Then \begin{equation} P_{j} = \sum_{j=1}^{J} w\, I(t_j)\, \delta t, \label{eq:pen_importance_nosub} \end{equation} \noindent where \begin{itemize}
  \item $\delta t$ --- time between measurements,
  \item $t_j$ --- time instant $j$,
  \item $w$ --- penalty rate per unit time (slice-level),
  \item $I(t_j)$ --- importance weight at time $t_j$ (dimensionless, typically $0<I(t_j)\le 1$),
  \item $T$ --- end time of a failure period,
  \item $\Delta T$ --- length of the failure period, and
  \item $J$ --- number of measurement intervals in failure period.
\end{itemize}

\subsubsection{Charge Dependent on the Time of Occurrence (Subcontracts)}
When subcontracts are taken into account, a \emph{subcontract} refers to an individual service component or metric commitment within the SLA (e.g., latency, throughput, or availability threshold). Let $k$ denote the number of violated subcontracts. For each subcontract $i$, let $\delta t_i$ denote the measurement interval, $w_i$ the penalty rate per unit time (at the subcontract level), $T_i$ the end time of a failure period, $\Delta T_i$ the duration of that failure period, and $I(i,t_j)$ the importance weight at time $t_j$ (dimensionless, typically $0 < I(i,t_j) \le 1$). Define $J_i:=\Delta T_i/\delta t_i$. Then \begin{equation} P_{ji} = \sum_{i=1}^{k} \sum_{j=1}^{J_i} w_i\, I(i,t_j)\, \delta t_i, \label{eq:pen_importance_sub} \end{equation} \noindent where \begin{itemize}
  \item $k$ --- number of exceeded subcontracts within the SLA,
  \item $\delta t_i$ --- time between measurements for subcontract $i$,
  \item $t_j$ --- time instant $j$,
  \item $w_i$ --- penalty rate per unit time for subcontract $i$ (subcontract-level),
  \item $I(i,t_j)$ --- importance weight of subcontract $i$ at time $t_j$ (dimensionless, typically $0<I(i,t_j)\le 1$),
  \item $T_i$ --- end time of a failure period for subcontract $i$,
  \item $\Delta T_i$ --- length of the failure period for subcontract $i$, and
  \item $J_i$ --- number of measurement intervals in the failure period of subcontract $i$.
\end{itemize}

\subsubsection{Total Penalty Within an Observation Window (Policy-based)}
To avoid double counting, the SLA policy should define the penalty granularity. In particular, penalties may be applied either at the slice level (without considering subcontracts) or at the subcontract level (with subcontracts), but not both simultaneously. Accordingly, the total penalty within the observation window can be defined as \begin{equation} P_{\mathrm{tot}} = P_{\mathrm{evt}} + \begin{cases}
P_t + P_{j}, & \text{\scriptsize(slice-level penalty, no subcontracts)}\\
P_{ji}, & \text{\scriptsize(subcontract-level penalty, weighted)}
\end{cases}
.
\label{eq:pen_total_policy}
\end{equation}

\subsection{Cost, Revenue, and Profit}
\vspace{-1.5mm}
The cost models of legacy telecommunication networks are typically based on capital expenditure (CAPEX) and operational expenditure (OPEX). In classical models, both CAPEX and OPEX are estimated according to factors such as traffic volume, the number of base stations, and energy consumption \cite{8024590}. However, this methodology is no longer appropriate for estimating the cost of slice-based 5G networks.

In sliced networks, network resources may be shared among multiple slices, and the network slicing scheme can vary across different resources. Therefore, OPEX cannot be estimated for the entire slice-based physical network as a single entity. Instead, a slice-oriented cost model is required to estimate the total cost, revenue, profit, and penalty associated with each individual slice, thereby helping to clearly define the SLA between the tenant and the provider.

As mentioned in Section~\ref{sec:Introduction}, each slice is designed to support a specific use case and has its own characteristics, QoS mechanisms, and architecture. Therefore, each slice must be identified by a subset of KPI requirements selected from a given set of KPIs $\mathbf{k} = [k_1, k_2, \ldots, k_L]$ through network function (NF) selection.

To estimate the required volume of network resources, we consider the NF implementation ($v$) and the size of the slice ($s$), where $s$ denotes the maximum number of user applications that can be served by the slice. Network resources may include various types, such as spectrum or bandwidth, power, time, human resources, and infrastructure. If the required amounts of these resources are represented by the vector $\mathbf{r} = [r_1, r_2, \ldots, r_N]$, where $N$ denotes the number of resource types, then by considering the cost associated with each resource, the resource requirements can be converted into the expenditure (EXP) in a manner similar to classical network cost models. Thus, we have \begin{equation}
\mathrm{EXP} = \mathrm{EXP}(\mathbf{r}),
\label{Service_EXP001}
\end{equation}

In practice, EXP(r) is computed by allocating per-slice resource usage over the SLA observation window and converting these quantities into monetary cost using operator-specific unit cost coefficients. \begin{equation}
\mathbf{r} = \mathbf{r}(\mathbf{k}, s, v).
\label{Service_EXP002}
\end{equation}

In practice, r(k,s,v) is obtained by mapping the slice KPI requirements k, slice size s, and the chosen NF implementation v to per-slice resource allocations using the operator’s internal KPI-to-resource planning and measurement data.

The tenant must pay a certain price for the service provided by the slice. Thus, given the service price ($p$), the slice size ($s$), and the customer size ($c$) (i.e., the number of user applications requesting service from the slice), the revenue (REV) generated by a slice can be modeled as \begin{equation}
\mathrm{REV} = \mathrm{REV}(p, s, c).
\label{Revenue_Slice}
\end{equation}

In practice, REV(p,s,c) is computed from the slice charging policy over the same observation window, using the agreed service price p and the measured customer demand c. To determine the profit generated by a slice, the cost is subtracted from the revenue, as follows: \begin{equation}
\pi = \mathrm{REV}(p, s, c) - \mathrm{EXP}(\mathbf{r}) = \pi(\mathbf{r}, p, s, c).
\label{Profit_Slice}
\end{equation}

It is important to note that the KPI-to-resource mapping described in Eq.~\ref{Service_EXP002} is highly complex and strongly dependent on the selection of the NF implementation ($v$). Nevertheless, since the network operator is responsible for the NF implementation, it possesses full knowledge of this mapping. Therefore, from the operator's perspective, it is reasonable to assume that the function $\mathbf{r}(\mathbf{k}, s, v)$ is \textit{a priori} known.

\subsection{QoS-related Metrics}
\vspace{-1.5mm}
The definitions and measurement units of QoS-related metrics, such as latency, delay, data rate, capacity, throughput, mobility, security, energy consumption, connection density, response time, and level of service -- are predefined by standardization organizations (e.g., ITU and ETSI). As widely discussed in the literature, slice-based 5G networks aim to achieve a 1000-fold increase in system capacity, a maximum data rate of 10~Gbps and an average individual user experience of 100~Mbps, prolonged battery life through a 1000-fold reduction in energy consumption per bit, a 90\% reduction in network energy usage, mobility support up to 500~km/h for high-speed users (e.g., high-speed trains), perceived availability of 99.99\%, 100\% coverage, and latency ranging from one millisecond to a few milliseconds \cite{HUAWEI001,Intel5G}.

Each slice is created based on a subset of these metrics to serve a specific number of users. Consequently, the business model, SLA structure, QoS specifications, and service levels may differ from slice to slice. While neither the tenant nor the operator can modify the definitions of these standardized metrics, the target values of these metrics can be adjusted within feasible limits through resource allocation and slice configuration. Therefore, both parties must specify the target values of these standardized metrics in SLA in accordance with the guidelines provided by standardization organizations.

\section{Conclusions} \label{sec:Conclusion}
\vspace{-1.5mm}
In this paper, we presented an E2E structure of an SLA between a tenant and an operator in 5G networks, aiming to balance the interests of both parties. The proposed SLA framework defines the reliability, availability, and performance of delivered telecommunication services to ensure that the right information reaches the right destination at the right time in a safe and secure manner. We also discussed key SLA metrics for slice-based networks, which are critical factors during the agreement process. In the future, we intend to investigate different types of SLAs, such as shared SLAs (where an SLA is shared among multiple tenants using the same slice) and hybrid SLAs (where certain tenants are prioritized while other authorized tenants may access the same slice). Moreover, this work can be extended by conducting a deeper analysis of additional QoS-related metrics, such as enhanced security mechanisms, reduced latency, and increased bandwidth.

%%%%%%%%%%%%%%%%%%%%%%%%%%%%%%%%%%%%%%%%%%%%%%%%%%%%%%%%%%%%%%%%%%%%%%%%%%%%%%%%%%%%%%%%%%%%%%%%%%%%%% Reference List %%%%%%%%%%%%%%%%%%%%%%%%%%%%%%%%%%%%%%%%%%%%%%%%%%%%%%%%%%%%%%%%%%%%%%%%%%%%%%%%%%%%%%%%%%%%%%%%%%

\bibliography{ref/mypaper01.bib} 
\bibliographystyle{ieeetr}
\vspace{-5.9mm}

%%%%%%%%%%%%%%%%%%%%%%%%%%%%%%%%%%%%%%%%%%%%%%%%%%%%%%%%%%%%%%%%%%%%%%%%%%%%%%%%%%%%%%%%%%%%%%%%%%%%%% Author Biography %%%%%%%%%%%%%%%%%%%%%%%%%%%%%%%%%%%%%%%%%%%%%%%%%%%%%%%%%%%%%%%%%%%%%%%%%%%%%%%%%%%%%%%%%%%%%%%%%%

\section*{Author Biographies}
\vskip 0pt plus -0.8fil

\begin{IEEEbiography}[{\includegraphics[width=1in,height=1.25in,clip,keepaspectratio]{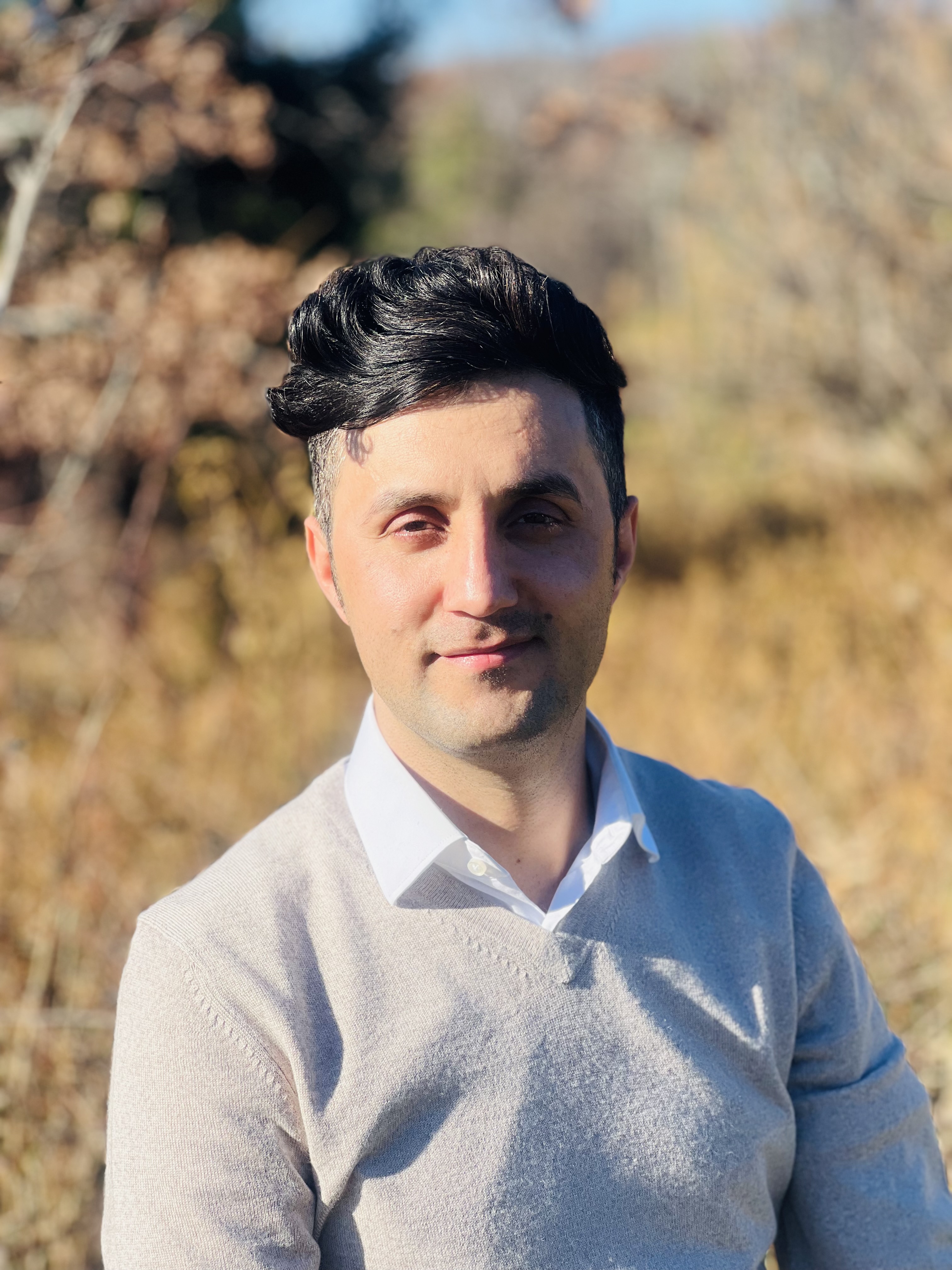}}]{Mohammad Asif Habibi} received his B.Sc. degree in telecommunication engineering from Kabul University, Afghanistan, in 2011. He obtained his M.Sc. degree in systems engineering and informatics from the Czech University of Life Sciences, Czech Republic, in 2016. Since January 2017, he has been working as a research fellow and Ph.D. candidate at the Division of Wireless Communications and Radio Navigation, Technische Universit\"at Kaiserslautern, Germany.  From 2011 to 2014, he worked as a radio access network engineer for HUAWEI. His main research interests include network slicing and radio access network.
\end{IEEEbiography} 
\vskip 0pt plus -1fil

\begin{IEEEbiography}[{\includegraphics[width=1in,height=1.25in,clip,keepaspectratio]{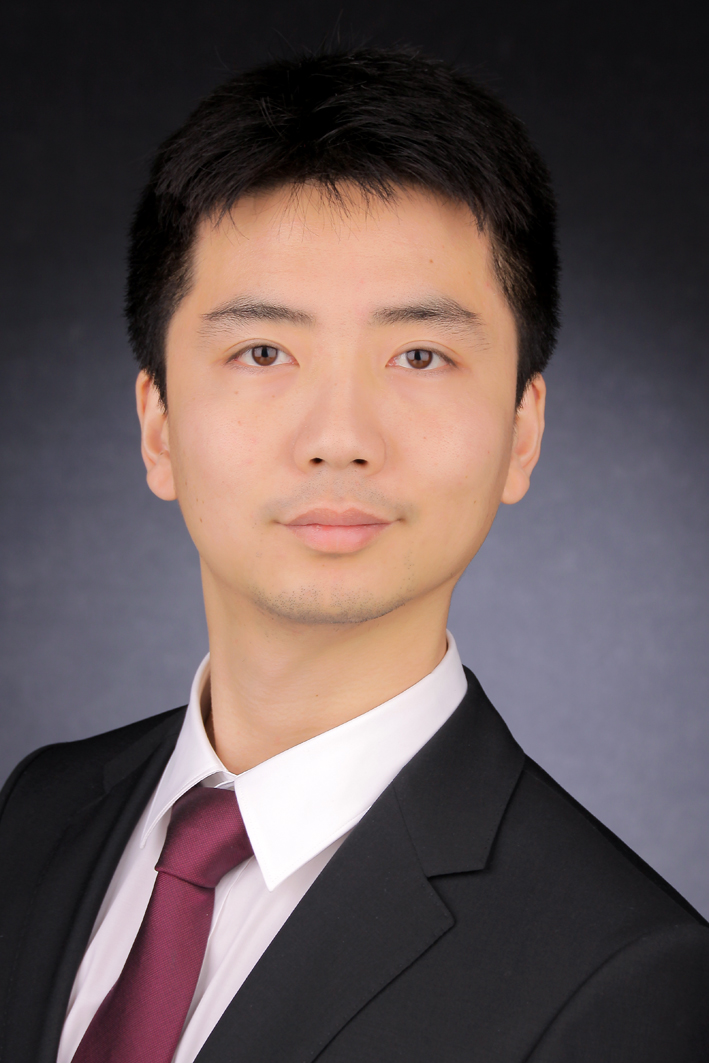}}]{Bin Han} (M'15 -- SM'21 ) received in 2009 his B.E. degree from Shanghai Jiao Tong University, in 2012 the M.Sc. from Technische Universit\"at Darmstadt, and in 2016 the Ph.D. (Dr.-Ing.) degree from Karlsruher Institute f\"ur Technologie. He joined Rheinland-Pf\"alzische Technische Universit\"at (previously known as Technische Universit\"at Kaiserslautern) in 2016, and is now a Senior Lecturer at its Division of Wireless Communications and Radio Navigation. %He has authored over around 50 research papers and book chapters.
\end{IEEEbiography} 
\vskip 0pt plus -1fil

\begin{IEEEbiography}[{\includegraphics[width=1in,height=1.25in,clip,keepaspectratio]{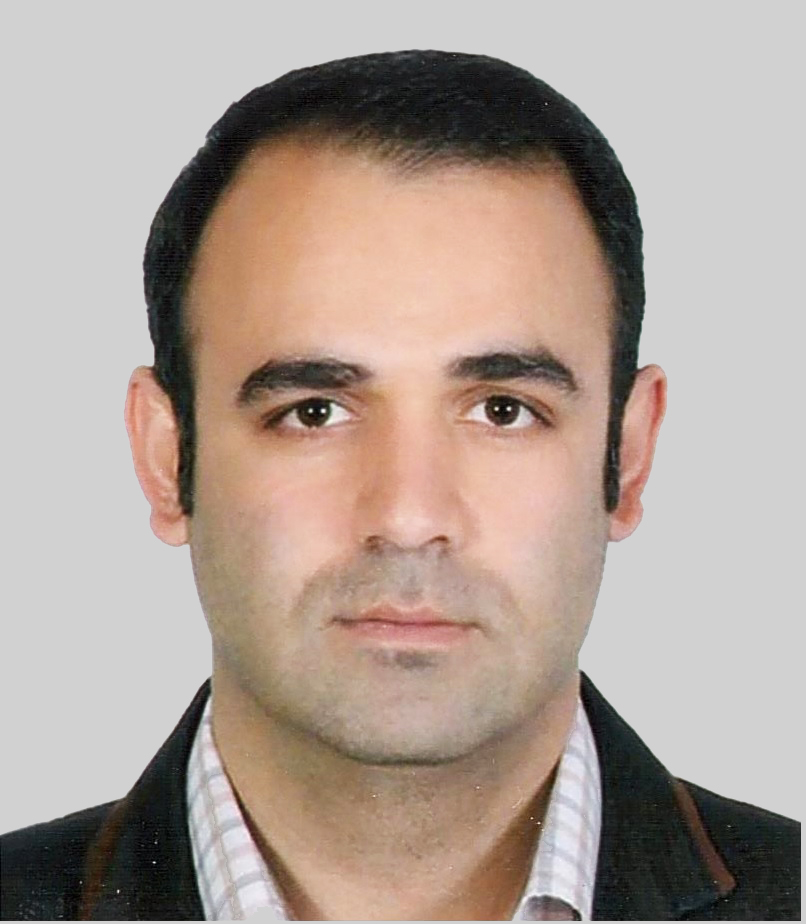}}]{Meysam Nasimi} received his M.Sc. degree in Communication Engineering from University Putra Malaysia (UPM) in 2014. Since 2017, he has been working as a Marie Curie Early Stage Researcher and Ph.D. candidate at the Institute of Wireless Communication, Technische Universit\"at Kaiserslautern, Germany. His current research interests are in the broad area of wireless communication systems, mobile edge computing and edge caching.
\end{IEEEbiography} 
\vskip 0pt plus -1fil

\begin{IEEEbiography}[{\includegraphics[width=1in,height=1.25in,clip,keepaspectratio]{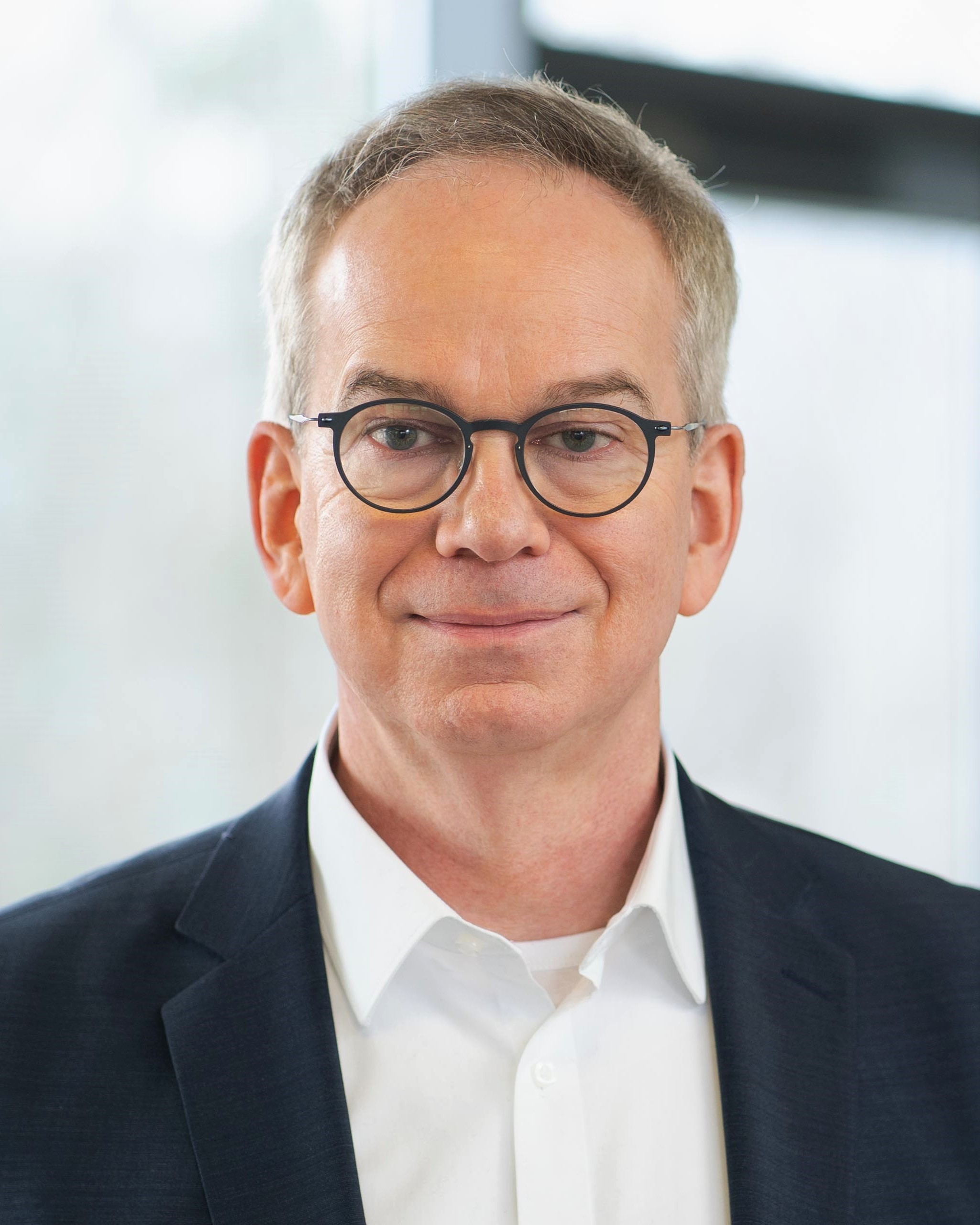}}]{Hans D. Schotten} received the Diploma and Ph.D. degrees in electrical engineering from the Aachen University of Technology, Germany, in 1990 and 1997, respectively. Since August 2007, he has been a full professor and head of the Division of Wireless Communications and Radio Navigation at Technische Universit\"at Kaiserslautern. Since 2012, he has also been Scientific Director at the German Research Center for Artificial Intelligence, heading the Intelligent Networks department. He was a senior researcher, the project manager, and the head of the research groups at Aachen University of Technology, Ericsson Corporate Research, and Qualcomm Corporate R\&D. %During his time at Qualcomm, he has also been the Director for Technical Standards and Coordinator of Qualcomm’s activities in European research programs.
\end{IEEEbiography}

\end{document}